\documentclass[12pt]{iopart}

\usepackage{epsfig,graphics}
\begin{document}
\title[Neutrino cooling rates due to $^{56}$Ni] {Expanded calculation of weak-interaction mediated neutrino cooling rates due
to $^{56}$Ni in stellar matter}
\author{Jameel-Un Nabi}
\address{Faculty of Engineering Sciences, GIK Institute of Engineering Sciences and
Technology, Topi 23640, NWFP, Pakistan } \ead{jnabi00@gmail.com}
\begin{abstract}
Accurate estimate of the neutrino cooling rates is required in order
to study the various stages of stellar evolution of massive stars.
Neutrino losses from proto-neutron stars play a crucial role in
deciding whether these stars would be crushed into black holes or
explode as supernovae. Both pure leptonic and weak-interaction
processes contribute to the neutrino energy losses in stellar
matter. At low temperatures and densities, characteristic of the
early phase of presupernova evolution, cooling through neutrinos
produced via the weak-interaction is important. Proton-neutron
quasi-particle random phase approximation (pn-QRPA) theory has
recently being used for calculation of stellar weak-interaction
rates of $fp$-shell nuclide with success. The lepton-to-baryon ratio
($Y_{e}$) during early phases of stellar evolution of massive stars
changes substantially alone due to electron captures on $^{56}$Ni.
The stellar matter is transparent to the neutrinos produced during
the presupernova evolution of massive stars. These neutrinos escape
the site and assist the stellar core in maintaining a lower entropy.
Here I present the expanded calculation of weak-interaction mediated
neutrino and antineutrino cooling rates due to $^{56}$Ni in stellar
matter using the pn-QRPA theory. This detailed scale is appropriate
for interpolation purposes and of greater utility for simulation
codes. The calculated rates are compared against earlier
calculations. During the relevant temperature and density regions of
stellar matter the reported rates show little differences with the
shell model rates and might contribute in fine-tuning of the
lepton-to-baryon ratio during the presupernova phases of stellar
evolution of massive stars.
\end{abstract}
 \maketitle

\section{Introduction}
It was the genius of Baade $\&$ Zwicky \cite{Baa34} who were able to
deduce the total energy released in a supernova explosion to be of
the order of ($3 \times 10^{51} - 10^{55}$) erg on the basis of a
few points of the light curve without any spectral information
available at that time. Later Colgate $\&$ White \cite{Col66} and
Arnett \cite{Arn67} presented their classical work on energy
transport by neutrinos and antineutrinos in non-rotating massive
stars. Since then we have come a long way and despite the immense
technological advancements the explosion mechanism of core-collapse
supernovae continues to pose challenges for the collapse simulators
throughout the globe. The prompt shock that follows the bounce
stagnates and is incapable of producing a supernova explosion on its
own. The stagnation is due to energy losses in disintegration of
iron nuclei (so far cooked in the stellar pot) and through neutrino
emissions (mainly non-thermal). The stellar matter is till then
transparent to the neutrinos emitted. A few milliseconds after the
bounce, the proto-neutron star accretes mass at a few tenths of
solar mass per second. This accretion, if continued even for one
second, can change the ultimate fate of the collapsing core
resulting into a black hole. Neutrinos have a crucial role to play
in this scenario and radiate around 10$\%$ of the rest mass
converting the star to a neutron star. Initially the nascent neutron
star is a hot thermal bath of dense nuclear matter, $e^{-}e^{+}$
pairs, photons and neutrinos. Neutrinos, having the
weak-interaction, are most effective in cooling and diffuse outward
within a few seconds, and eventually escape with about 99$\%$ of the
released gravitational energy. Despite the small neutrino-nucleus
cross sections, the neutrinos flux generated by the cooling of a
neutron star can produce a number of nuclear transmutations as it
passes the onion-like structured envelope surrounding the neutron
star. The microphysics involved in these extreme processes is indeed
complex and one should be very cautious in interpolating and/or
extrapolating values of stellar parameters during various phases of
stellar evolution. A lot many physical inputs are required at the
beginning of each stage of the entire simulation process (e.g.
collapse of the core, formation, stalling and revival of the shock
wave and shock propagation). It is highly desirable to calculate
these parameters with the most reliable physical data and inputs.

During the late phases of evolution of massive stars an iron core
develops (of mass around  $1.5 M_{\odot}$). Capture rates and
photodisintegration processes contribute in lowering of the
degeneracy pressure required to counter the enormous self-gravity
force of the star. Under such extreme thermodynamic conditions,
neutrinos are produced in abundance. Eventually the collapse of the
iron core begins. The mechanism of core-collapse supernovae is
strongly believed to depend upon the transfer of energy from the
inner core to the outer mantle of the iron core. Neutrinos seem to
be the mediators of this energy transfer. As mentioned above the
shock wave, produced as a result, stalls due to photodisintegration
and neutrino energy losses. Once again the part played by neutrinos
in this scenario is far from being completely understood. In the
late-time neutrino heating mechanism the stalled shock can be
revived (about 1 s after the bounce) and may be driven as a delayed
explosion \cite{Bet85}. However to date there have been no
successfully simulated spherically symmetric explosions. Even the 2D
simulations (addition of convection) performed with a Boltzmann
solver for the neutrino transport fails to convert the collapse into
an explosion \cite{Bur03}. (Recently a few simulation groups (e.g.
\cite{Buras06, Bur06, Woo07} have reported successful explosions in
2D mode.) Additional energy sources (e.g. magnetic fields and
rotations) were also sought that might transport energy to the
mantle and lead to an explosion. World-wide core-collapse simulators
are still working hard to come up with a convincing and decisive
mode of producing explosions.

Neutrinos from core-collapse supernovae are unique messengers of the
microphysics of supernovae and are crucial to the life and afterlife
of supernovae. They provide information regarding the neutronization
due to electron capture, the infall phase, the formation and
propagation of the shock wave and the cooling phase. Cooling rate is
one of the crucial parameters that strongly affects the stellar
evolution. In stellar matter the neutrinos are produced from both
weak-interaction reactions and pure leptonic processes. The later
includes pair annihilation, bremsstrahlung on nuclei, plasmon decay
and $\nu$-photoproduction processes. White dwarfs and supernovae
(which are the endpoints for stars of varying masses) have both
cooling rates largely dominated by neutrino production. A cooling
proto-neutron star emits about $3 \times 10^{53}$ erg in neutrinos,
with the energy roughly equipartitioned among all species. The
neutrino energy loss rates are important input parameters in
multi-dimensional simulations of the contracting proto-neutron star.
Parameter-free multi-dimensional models, with neutrino transport
included consistently throughout the entire mass, yield conflicting
results on the key issue of whether the star actually explodes.
Reliable and microscopic calculations of neutrino loss rates and
capture rates can contribute effectively in the final outcome of
these simulations. Whereas neutrinos produced via pure leptonic
processes dominate in the very high temperature-density domain
during the very late phases of stellar evolution, the
weak-interaction neutrinos also play an important role in cooling
the core to a lower entropy specially during the early phases of
stellar evolution. This work is primarily devoted to calculate the
neutrino energy loss rates due to weak-interaction reactions
(capture and $\beta$ decays) on $^{56}$Ni.

$^{56}$Ni is abundant in the presupernova conditions and
weak-interaction reactions on this nucleus is believed to contribute
effectively in the dynamics of presupernova evolution. Aufderheide
and collaborators \cite{Auf94} ranked $^{56}$Ni as the third most
important electron capture nucleus averaged throughout the stellar
trajectory for 0.4 $\le Y_{e} \le$ 0.5 during the presupernova
evolution. Later Heger et al. \cite{Heg01} also identified $^{56}$Ni
as one of the most important nuclide for capture purposes for the
presupernova evolution of massive stars ($25M_{\odot}$ and
$40M_{\odot}$). Realizing the importance of $^{56}$Ni in
astrophysical environments, Nabi and Rahman \cite{Nab05} reported
the calculation of electron capture rates on $^{56}$Ni using the
pn-QRPA theory (see also Ref. \cite{Nab08} regarding the calculation
of ground and excited state Gamow-Teller (GT) strength distributions
of $^{56}$Ni). In a recent review of theory of core-collapse
supernovae, Janka and collaborators \cite{Jan07} again discussed the
importance of electron capture rates on $^{56}$Ni in presupernova
evolution of massive stars. It can be seen from Fig. 1 (Ref.
\cite{Jan07}) that during the onset of collapse (where t $\sim$ 0s)
to the later stages (t $\sim$ 0.11s), iron and nickel isotopes are
present in reasonable quantity. According to Auderheide and
collaborators \cite{Auf94}, for $Y_{e}$ around 0.5, $^{56}$Ni is the
most abundant nucleus having a mass fraction of around 0.99.
Consequently electron capture rates on $^{56}$Ni is an important
process during these phases. During later stages after the bounce
and shock propagation (t $\sim$ 0.12s) the photodisintegration of
iron-group nuclei to alpha particles and protons results which marks
the beginning of the proto-neutron star.

In this paper I analyze the weak-interaction neutrino energy loss
rates (which I term as neutrino cooling rates throughout this text)
due to this key isotope of nickel which is so abundant during the
silicon burning phases of the stellar core. Due to the extreme
conditions prevailing in these scenarios, interpolation of
calculated rates within large intervals of temperature-density
points might pose some uncertainty in the values of weak rates for
collapse simulators. In this paper I describe the calculation of the
neutrino and antineutrino cooling rates due to capture and decay
rates on $^{56}$Ni on an expanded temperature-density grid suitable
for collapse simulation codes. Section 2 briefly discusses the
formalism of the pn-QRPA calculations and presents some of the
calculated results. Comparison with earlier calculations during
stellar evolution of massive stars is also included in this section.
I summarize the main conclusions in Section 3 and at the end Table I
presents the expanded calculation of neutrino and antineutrino
cooling rates due to $^{56}$Ni in stellar matter.

\section{Calculations and Results}
The Hamiltonian of the pn-QRPA model, the model parameters and their
selection criteria were discussed earlier in Ref. \cite{Nab08}. The
neutrino (antineutrino) cooling rates can occur through four
different weak-interaction mediated channels: electron and positron
emissions, and, continuum electron and positron captures. It is
assumed that the stellar matter is transparent to the neutrinos and
antineutrinos produced as a result of these reactions during the
presupernova evolutionary phases and contributes effectively in
cooling the system. The neutrino (antineutrino) cooling rates were
calculated using the relation

\begin{equation}
\lambda ^{^{\nu(\bar{\nu})} } _{ij} =\left[\frac{\ln 2}{D}
\right]\left[f_{ij}^{\nu} (T,\rho ,E_{f} )\right]\left[B(F)_{ij}
+\left({\raise0.7ex\hbox{$ g_{A}  $}\!\mathord{\left/ {\vphantom
{g_{A}  g_{V} }} \right.
\kern-\nulldelimiterspace}\!\lower0.7ex\hbox{$ g_{V}  $}}
\right)^{2} B(GT)_{ij} \right]. \label{wi}
\end{equation}
The value of D was taken to be 6295s \cite{Yos88}. $B_{ij}'s$ are
the sum of reduced transition probabilities of the Fermi B(F) and GT
transitions B(GT). The effective ratio of axial and vector coupling
constants, $(g_{A}/g_{V})$, was taken to be -1.254 \cite{Rod06}. The
$f_{ij}^{\nu}$ are the phase space integrals and are functions of
stellar temperature ($T$), density ($\rho$) and Fermi energy
($E_{f}$) of the electrons. They are explicitly given by
\begin{equation}
f_{ij}^{\nu} \, =\, \int _{1 }^{w_{m}}w\sqrt{w^{2} -1} (w_{m} \,
 -\, w)^{3} F(\pm Z,w)(1- G_{\mp}) dw,
\label{phdecay}
\end{equation}
and by
\begin{equation}
f_{ij}^{\nu} \, =\, \int _{w_{l} }^{\infty }w\sqrt{w^{2} -1} (w_{m}
\,
 +\, w)^{3} F(\pm Z,w)G_{\mp} dw.
\label{phcapture}
\end{equation}
In Eqs. ~(\ref{phdecay}) and ~(\ref{phcapture}),  $w$ is the total
energy of the electron including its rest mass, $w_{l}$ is the total
capture threshold energy (rest+kinetic) for positron (or electron)
capture. F($\pm$ Z,w) are the Fermi functions and were calculated
according to the procedure adopted by Gove and Martin \cite{Gov71}.
G$_{\pm}$ is the Fermi-Dirac distribution function for positrons
(electrons).
\begin{equation}
G_{+} =\left[\exp \left(\frac{E+2+E_{f} }{kT}\right)+1\right]^{-1},
\label{Gp}
\end{equation}
\begin{equation}
 G_{-} =\left[\exp \left(\frac{E-E_{f} }{kT}
 \right)+1\right]^{-1},
\label{Gm}
\end{equation}
here $E$ is the kinetic energy of the electrons and $k$ is the
Boltzmann constant.

For the decay (capture) channel Eq. ~(\ref{phdecay}) (Eq.
~(\ref{phcapture})) was used for the calculation of phase space
integrals. Upper signs were used for the case of electron emissions
(captures) and lower signs for the case of positron emissions
(captures). Details of the calculation of reduced transition
probabilities can be found in Ref. \cite{Nab04}.

The construction of parent and daughter excited states and
calculation of GT transition amplitudes connecting these states
within the pn-QRPA model is very important. The excited states in
the pn-QRPA model can be constructed as phonon-correlated
multi-quasi-particle states. The RPA is formulated for excitations
from the $J^{\pi} = 0^{+}$ ground state of an even-even nucleus. The
model extended to include the quasiparticle transition degrees of
freedom yields decay half-lives of odd-mass and odd-odd parent
nuclei with the same quality of agreement with experiment as for
even-even nuclei (where only QRPA phonons contribute to the decays)
\cite{Mut92}. When the parent nucleus has an odd nucleon, the ground
state can be expressed as a one-quasiparticle state, in which the
odd quasiparticle (q.p) occupies the single-q.p. orbit of the
smallest energy.  Then there exits two different type of
transitions: phonon transitions with the odd nucleon acting only as
a spectator and transition of the odd nucleon itself. For the later
case, phonon correlations were introduced to one-quasiparticle
states in first-order perturbation \cite{Mut89}. The transition
amplitudes between the multi-quasi-particle states can be reduced to
those of single-quasi-particle states as shown below.

Excited states of $^{56}$Ni can be constructed as two-proton
quasiparticle (q.p) states and two-neutron q.p. states. Transitions
from these initial states are possible to final proton-neutron q.p.
pair states in the odd-odd daughter nucleus. The transition
amplitudes and their reduction to correlated ($c$) one-q.p.states
are given by
\begin{eqnarray}
<p^{f}n_{c}^f \mid t_{\pm}\sigma_{-\mu} \mid p_{1}^{i}p_{2c}^{i}> \nonumber \\
 = -\delta (p^{f},p_{2}^{i}) <n_{c}^{f} \mid t_{\pm}\sigma_{-\mu} \mid p_{1c}^{i}>
+\delta (p^{f},p_{1}^{i}) <n_{c}^{f} \mid t_{\pm}\sigma_{-\mu} \mid
p_{2c}^{i}>
 \label{first}
\end{eqnarray}
\begin{eqnarray}
<p^{f}n_{c}^f \mid t_{\pm}\sigma_{\mu} \mid n_{1}^{i}n_{2c}^{i}> \nonumber \\
 = +\delta (n^{f},n_{2}^{i}) <p_{c}^{f} \mid t_{\pm}\sigma_{\mu} \mid n_{1c}^{i}>
-\delta (n^{f},n_{1}^{i}) <p_{c}^{f} \mid t_{\pm}\sigma_{\mu} \mid
n_{2c}^{i}>
\end{eqnarray}
where $\mu$ = -1, 0, 1, are the spherical components of the spin
operator and $t_{\pm}$ is the isospin raising and lowering operator.

For an odd-odd nucleus the ground state is assumed to be a
proton-neutron q.p. pair state of smallest energy. States in
$^{56}$Co, $^{56}$Cu are expressed in q.p. transformation by
two-q.p. states (proton-neutron pair states) or by four-q.p. states
(two-proton or two-neutron q.p. states). Reduction of two-q.p.
states into correlated ($c$) one-q.p. states is given as
\begin{eqnarray}
<p_{1}^{f}p_{2c}^{f} \mid t_{\pm}\sigma_{\mu} \mid p^{i}n_{c}^{i}> \nonumber\\
= \delta(p_{1}^{f},p^{i}) <p_{2c}^{f} \mid t_{\pm}\sigma_{\mu} \mid
n_{c}^{i}> - \delta(p_{2}^{f},p^{i}) <p_{1c}^{f} \mid
t_{\pm}\sigma_{\mu} \mid n_{c}^{i}>
\end{eqnarray}
\begin{eqnarray}
<n_{1}^{f}n_{2c}^{f} \mid t_{\pm}\sigma_{-\mu} \mid p^{i}n_{c}^{i}> \nonumber\\
= \delta(n_{2}^{f},n^{i}) <n_{1c}^{f} \mid t_{\pm}\sigma_{-\mu} \mid
p_{c}^{i}> - \delta(n_{1}^{f},n^{i}) <n_{2c}^{f} \mid
t_{\pm}\sigma_{-\mu} \mid p_{c}^{i}>
\end{eqnarray}
while the four-q.p. states are simplified as
\begin{eqnarray}
<p_{1}^{f}p_{2}^{f}n_{1}^{f}n_{2c}^{f} \mid t_{\pm}\sigma_{-\mu}
\mid p_{1}^{i}p_{2}^{i}p_{3}^{i}n_{1c}^{i}>
\nonumber\\
=\delta (n_{2}^{f},n_{1}^{i})[ \delta (p_{1}^{f},p_{2}^{i})\delta
(p_{2}^{f},p_{3}^{i})
<n_{1c}^{f} \mid t_{\pm}\sigma_{-\mu} \mid p_{1c}^{i}> \nonumber\\
-\delta (p_{1}^{f},p_{1}^{i}) \delta (p_{2}^{f},p_{3}^{i})
<n_{1c}^{f} \mid t_{\pm}\sigma_{-\mu} \mid p_{2c}^{i}> +\delta
(p_{1}^{f},p_{1}^{i}) \delta (p_{2}^{f},p_{2}^{i}) \nonumber\\
<n_{1c}^{f} \mid t_{\pm}\sigma_{-\mu} \mid p_{3c}^{i}>]  -\delta
(n_{1}^{f},n_{1}^{i})[ \delta (p_{1}^{f},p_{2}^{i})\delta
(p_{2}^{f},p_{3}^{i})
<n_{2c}^{f} \mid t_{\pm}\sigma_{-\mu} \mid p_{1c}^{i}> \nonumber\\
-\delta (p_{1}^{f},p_{1}^{i}) \delta (p_{2}^{f},p_{3}^{i})
<n_{2c}^{f} \mid t_{\pm}\sigma_{-\mu} \mid p_{2c}^{i}> +\delta
(p_{1}^{f},p_{1}^{i}) \delta (p_{2}^{f},p_{2}^{i})\nonumber\\
<n_{2c}^{f}\mid t_{\pm}\sigma_{-\mu} \mid p_{3c}^{i}>]
\end{eqnarray}
\begin{eqnarray}
<p_{1}^{f}p_{2}^{f}p_{3}^{f}p_{4c}^{f} \mid t_{\pm}\sigma_{\mu} \mid
p_{1}^{i}p_{2}^{i}p_{3}^{i}n_{1c}^{i}>
\nonumber\\
=-\delta (p_{2}^{f},p_{1}^{i}) \delta (p_{3}^{f},p_{2}^{i})\delta
(p_{4}^{f},p_{3}^{i})
<p_{1c}^{f} \mid t_{\pm}\sigma_{\mu} \mid n_{1c}^{i}> \nonumber\\
+\delta (p_{1}^{f},p_{1}^{i}) \delta (p_{3}^{f},p_{2}^{i}) \delta
(p_{4}^{f},p_{3}^{i})
<p_{2c}^{f} \mid t_{\pm}\sigma_{\mu} \mid n_{1c}^{i}> \nonumber\\
-\delta (p_{1}^{f},p_{1}^{i}) \delta (p_{2}^{f},p_{2}^{i}) \delta
(p_{4}^{f},p_{3}^{i})
<p_{3c}^{f} \mid t_{\pm}\sigma_{\mu} \mid n_{1c}^{i}> \nonumber\\
+\delta (p_{1}^{f},p_{1}^{i}) \delta (p_{2}^{f},p_{2}^{i}) \delta
(p_{3}^{f},p_{3}^{i}) <p_{4c}^{f} \mid t_{\pm}\sigma_{\mu} \mid
n_{1c}^{i}>
\end{eqnarray}
\begin{eqnarray}
<p_{1}^{f}p_{2}^{f}n_{1}^{f}n_{2c}^{f} \mid t_{\pm}\sigma_{\mu} \mid
p_{1}^{i}n_{1}^{i}n_{2}^{i}n_{3c}^{i}>
\nonumber\\
=\delta (p_{1}^{f},p_{1}^{i})[ \delta (n_{1}^{f},n_{2}^{i})\delta
(n_{2}^{f},n_{3}^{i})
<p_{2c}^{f} \mid t_{\pm}\sigma_{\mu} \mid n_{1c}^{i}> \nonumber\\
-\delta (n_{1}^{f},n_{1}^{i}) \delta (n_{2}^{f},n_{3}^{i})
<p_{2c}^{f} \mid t_{\pm}\sigma_{\mu} \mid n_{2c}^{i}> +\delta
(n_{1}^{f},n_{1}^{i}) \delta (n_{2}^{f},n_{2}^{i})\nonumber\\
<p_{2c}^{f} \mid t_{\pm}\sigma_{\mu} \mid n_{3c}^{i}>]  -\delta
(p_{2}^{f},p_{1}^{i})[ \delta (n_{1}^{f},n_{2}^{i})\delta
(n_{2}^{f},n_{3}^{i})
<p_{1c}^{f} \mid t_{\pm}\sigma_{\mu} \mid n_{1c}^{i}> \nonumber\\
-\delta (n_{1}^{f},n_{1}^{i}) \delta (n_{2}^{f},n_{3}^{i})
<p_{1c}^{f} \mid t_{\pm}\sigma_{\mu} \mid n_{2c}^{i}> +\delta
(n_{1}^{f},n_{1}^{i}) \delta (n_{2}^{f},n_{2}^{i}) \nonumber\\
<p_{1c}^{f} \mid t_{\pm}\sigma_{\mu} \mid n_{3c}^{i}>]
\end{eqnarray}
\begin{eqnarray}
<n_{1}^{f}n_{2}^{f}n_{3}^{f}n_{4c}^{f} \mid t_{\pm}\sigma_{-\mu}
\mid p_{1}^{i}n_{1}^{i}n_{2}^{i}n_{3c}^{i}>
\nonumber\\
= +\delta (n_{2}^{f},n_{1}^{i}) \delta (n_{3}^{f},n_{2}^{i})\delta
(n_{4}^{f},n_{3}^{i})
<n_{1c}^{f} \mid t_{\pm}\sigma_{-\mu} \mid p_{1c}^{i}> \nonumber\\
-\delta (n_{1}^{f},n_{1}^{i}) \delta (n_{3}^{f},n_{2}^{i}) \delta
(n_{4}^{f},n_{3}^{i})
<n_{2c}^{f} \mid t_{\pm}\sigma_{-\mu} \mid p_{1c}^{i}> \nonumber\\
+\delta (n_{1}^{f},n_{1}^{i}) \delta (n_{2}^{f},n_{2}^{i}) \delta
(n_{4}^{f},n_{3}^{i})
<n_{3c}^{f} \mid t_{\pm}\sigma_{-\mu} \mid p_{1c}^{i}> \nonumber\\
-\delta (n_{1}^{f},n_{1}^{i}) \delta (n_{2}^{f},n_{2}^{i}) \delta
(n_{3}^{f},n_{3}^{i}) <n_{4c}^{f} \mid t_{\pm}\sigma_{-\mu} \mid
p_{1c}^{i}> \label{last}
\end{eqnarray}
For all the given q.p. transition amplitudes [Eqs. ~(\ref{first})-
~(\ref{last})],
the antisymmetrization of the single- q.p. states was taken into account:\\
$ p_{1}^{f}<p_{2}^{f}<p_{3}^{f}<p_{4}^{f}$,\\
$ n_{1}^{f}<n_{2}^{f}<n_{3}^{f}<n_{4}^{f}$,\\
$ p_{1}^{i}<p_{2}^{i}<p_{3}^{i}<p_{4}^{i}$,\\
$ n_{1}^{i}<n_{2}^{i}<n_{3}^{i}<n_{4}^{i}$.\\
GT transitions of phonon excitations for every excited state were
also taken into account. Here I assumed that the quasiparticles in
the parent nucleus remained in the same quasiparticle orbits. A
detailed description of the formalism for the calculation of GT
transition amplitudes can be found in Ref. \cite{Mut92}.

The total neutrino cooling rate per unit time per nucleus is given
by
\begin{equation}
\lambda^{\nu} =\sum _{ij}P_{i} \lambda _{ij}^{\nu},
\label{nurate}
\end{equation}
where $\lambda_{ij}^{\nu}$ is the sum of the electron capture and
positron decay rates for the transition $i \rightarrow j$ and
$P_{i}$ is the probability of occupation of parent excited states
which follows the normal Boltzmann distribution.

On the other hand the total antineutrino cooling rate per unit time
per nucleus is given by
\begin{equation}
\lambda^{\bar{\nu}} =\sum _{ij}P_{i} \lambda _{ij}^{\bar{\nu}},
\label{nubarrate}
\end{equation}
where $\lambda_{ij}^{\bar{\nu}}$ is the sum of the positron capture
and electron decay rates for the transition $i \rightarrow j$.

The summation over all initial and final states was carried out
until satisfactory convergence in the rate calculation was achieved.
The pn-QRPA theory allows a microscopic state-by-state calculation
of both sums present in Eqs. ~(\ref{nurate}) and ~(\ref{nubarrate}).
This feature of the pn-QRPA model greatly increases the reliability
of the calculated rates over other models in stellar matter where
there exists a finite probability of occupation of excited states.

\begin{figure}[htbp]
\begin{center}
\includegraphics[width=0.8\textwidth]{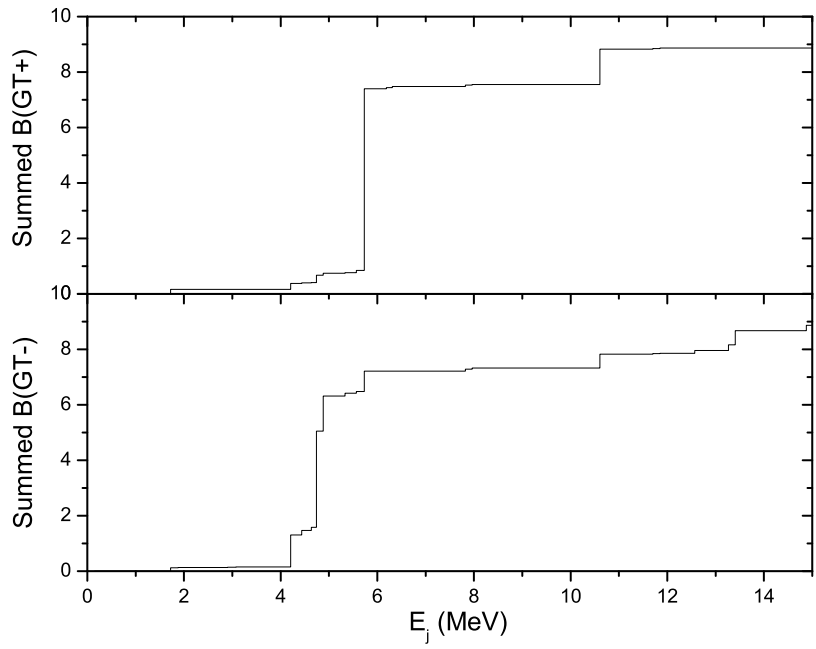}
\caption{Cumulative sum of the calculated B(GT${\pm}$) values.
$E_{j}$ represents the excitation energies in daughter
nuclei.}\label{figure1}
\end{center}
\end{figure}

Experimental data were incorporated wherever available to strengthen
the reliability of the calculation. The calculated excitation
energies (along with their $logft$ values) were replaced with the
measured ones when they were within 0.5 MeV of each other. Missing
measured states were inserted and inverse and mirror transitions
were also taken into consideration. If there appeared a level in
experimental compilations without definite spin and/or parity
assignment, theoretical levels were not replaced (inserted) with the
experimental ones beyond this excitation energy. The detailed
analysis of the pn-QRPA calculated ground and excited state
GT$_{\pm}$ strength distributions of $^{56}$Ni was presented earlier
in Ref. \cite{Nab08}. The pn-QRPA model calculated the centroid of
the GT$_{+}$ strength distribution to be around 5.7 MeV. This is to
be compared with the FFN \cite{Ful82} value  of 3.8 MeV and large
scale shell model range of 2.5 -- 3.0 MeV \cite{Lan98}. Here I
present the ground state cumulative GT strength in both directions
($\Sigma S_{\beta^{\pm}}$) for $^{56}$Ni in Figure~\ref{figure1}. In
the figure the upper panel shows the calculated summed B(GT$_{+}$)
strength distribution ($\Sigma S_{\beta^{+}}$) whereas the bottom
panel depicts the calculated summed B(GT$_{-}$) strength
distribution ($\Sigma S_{\beta^{-}}$). The abscissa shows the
daughter excitation energy ($^{56}$Co in upper panel and $^{56}$Cu
in lower panel) in units of MeV. It can be seen from
Figure~\ref{figure1} that the GT$_{\pm}$ distributions are
well-fragmented and extend to high-lying daughter states. The model
independent Ikeda sum rule is fulfilled in the calculation.

\begin{figure}[htbp]
\begin{center}
\includegraphics[width=0.8\textwidth]{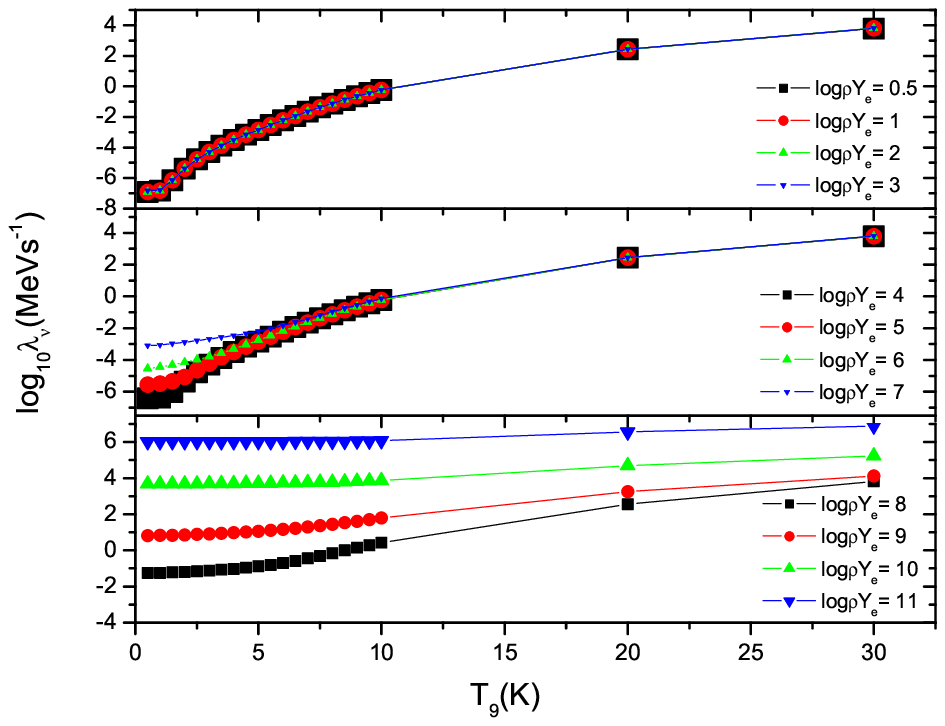}
\caption{(Color online) Neutrino cooling rates due to $^{56}$Ni, as
function of temperature, for different selected densities .
Densities are in units of $gcm^{-3}$, $T_{9}$ represents
temperatures in $10^{9}$ K and log$_{10}\lambda_{\nu}$ represents
the log of neutrino cooling rates. }\label{figure2}
\end{center}
\end{figure}

The pn-QRPA calculated neutrino cooling rates are depicted in
Figure~\ref{figure2}. The figure shows the calculated rates as a
function of stellar temperatures and densities. The upper, middle
and lower panels depict the cooling rates at low ($\rho Y_{e}
[gcm^{-3}] =10^{0.5}, 10^{1}$, $10^{2}$ and $10^{3}$),  medium
($\rho Y_{e} [gcm^{-3}] =10^{4}, 10^{5}$, $10^{6}$ and $10^{7}$),
and high ($\rho Y_{e} [gcm^{-3}] =10^{8}, 10^{9}$, $10^{10}$ and
$10^{11}$) stellar densities, respectively. The neutrino energy loss
rates are given in logarithmic scales (to base 10) in units of $MeV.
s^{-1}$. In the figures and throughout the text T$_{9}$ gives the
stellar temperature in units of $10^{9}$ K. It can be seen from
Figure~\ref{figure2} that at low stellar densities the cooling rates
remain more or less the same as one increases the density by an
order of magnitude. Considerable enhancement in neutrino cooling
rates is witnessed as the stellar cores attain medium and high
densities. This difference is more prominent in low temperature
domain T$_{9}$ $<$ 5. Specially in high density region of stellar
core, the neutrino cooling rates increase by orders of magnitude as
the core stiffens further. For a given temperature the neutrino
energy loss rates increase monotonically with increasing densities.

\begin{figure}[htbp]
\begin{center}
\includegraphics[width=0.8\textwidth]{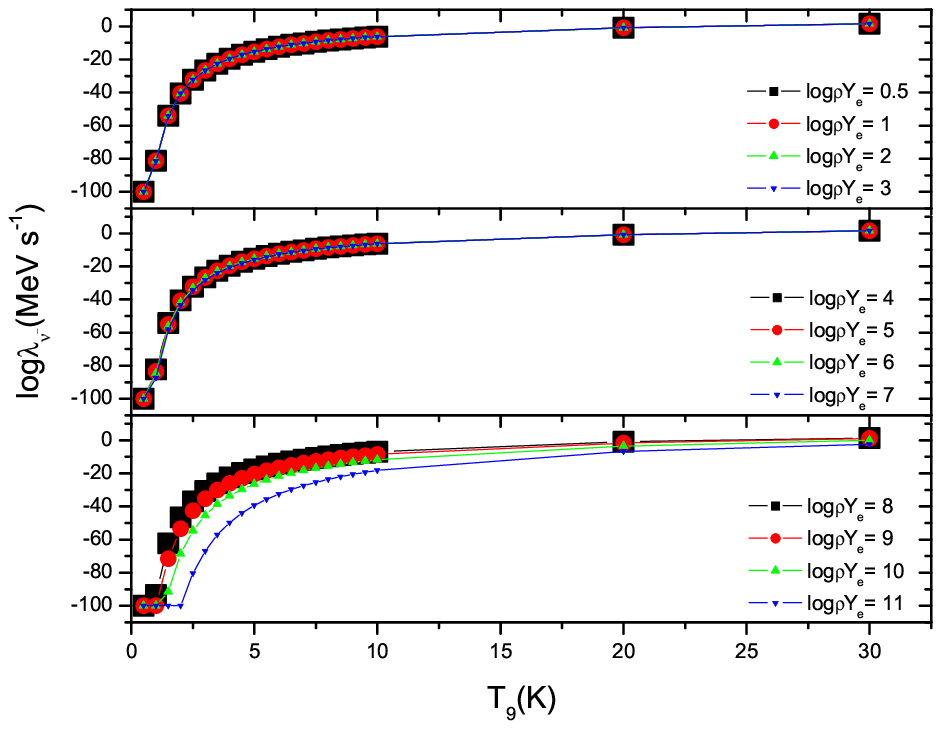}
\caption{(Color online) Antineutrino cooling rates due to $^{56}$Ni,
as function of temperature, for different selected densities .
Densities are in units of $gcm^{-3}$, $T_{9}$ represents
temperatures in $10^{9}$ K and and log$_{10}\lambda_{\bar{\nu}}$
represents the log of antineutrino cooling rates.}\label{figure3}
\end{center}
\end{figure}

The antineutrino energy loss rates are very small in magnitude as
compared to the neutrino energy loss rates and as such these rates
have a very small contribution in cooling of the stellar cores. This
is because the positron capture on $^{56}$Ni as well as the
$\beta$-decay of $^{56}$Ni is relatively suppressed as compared to
the electron capture rates on $^{56}$Ni. The calculated antineutrino
cooling rates are depicted in Figure~\ref{figure3}. Once again I
show the result in a three-panel format as before. There is a sharp
exponential increase in the antineutrino cooling rates as the
stellar temperature increases up to T$_{9}$ =5. Beyond this
temperature the slope of the rates reduces with increasing density.
For a given temperature the antineutrino energy loss rates increase
monotonically with increasing densities. The rates are almost
superimposed on one another as a function of stellar densities in
low density domain. However as the stellar matter moves from the
medium high density region to high density region these rates start
to 'peel off' from one another. The neutrino and antineutrino energy
loss rates are calculated on an extensive temperature-density grid
point suitable for collapse simulations and interpolation purposes
and presented at the end of this paper in Table I. The calculated
rates are tabulated in logarithmic (to base 10) scale. In the table,
-100 means that the rate is smaller than 10$^{-100} MeV. s^{-1}$ .
The first column gives log($\rho Y_{e}$) in units of $g cm^{-3}$,
where $\rho$ is the baryon density and $Y_{e}$ is the ratio of the
electron number to the baryon number. Stellar temperatures ($T_{9}$)
are stated in $10^{9} K$. Stated also are the values of the Fermi
energy of electrons in units of $MeV$.
$\lambda_{\bar{\nu}}$($\lambda_{\bar{\nu}}$) are the
neutrino(antineutrino) cooling rates in units of  $MeV. s^{-1}$. The
electronic versions (ASCII files) of these rates may be requested
from the author.

The calculation of neutrino cooling rates was also compared with
previous calculations. For the sake of comparison I took into
consideration the pioneer calculations of FFN \cite{Ful82} and those
performed using the large-scale shell model (LSSM) \cite{Lan00}. The
FFN rates were used in many simulation codes (e.g. KEPLER stellar
evolution code \cite{Wea78}) while LSSM rates were employed in
recent simulation of presupernova evolution of massive stars in the
mass range 11-40 $M_{\odot}$ \cite{Heg01}. The neutrino energy loss
rates have contributions both from electron capture and positron
decay rates (Eq. ~(\ref{nurate})). Both of these weak interaction
mediated processes are governed by the ground and excited state
GT$_{+}$ strength distributions. As mentioned earlier the pn-QRPA
model places the centroid of the ground state GT$_{+}$ strength
distribution around 3 MeV (2 MeV) higher than the LSSM (FFN)
centroid. Accordingly one expects a somewhat larger neutrino cooling
rates due to previous calculations as compared to the reported
rates. However at low temperatures and densities the pn-QRPA
calculated positron decay rates are orders of magnitude bigger which
causes an overall enhancement of the neutrino cooling rates (around
a factor 5). During intermediate stellar temperature and density
domains the three calculations are in excellent agreement. At higher
temperatures and densities the pn-QRPA neutrino cooling rates are
smaller up to an order of magnitude. The detailed comparison is
presented below.

Figure~\ref{figure4} depicts the comparison of neutrino cooling
rates due to $^{56}$Ni with earlier calculations at low densities.
The upper panel displays the ratio of calculated rates to the LSSM
rates, $R_{\nu}(QRPA/LSSM)$, while the lower panel shows a similar
comparison with the FFN calculation, $R_{\nu}(QRPA/FFN)$. The graph
is drawn for the low-density regions ($\rho Y_{e} [gcm^{-3}]
=10^{1}, 10^{3}, 10^{5}$) as a function of stellar temperature. Both
graphs follow a similar trend. At low densities and temperatures the
pn-QRPA cooling rates are bigger by as much as a factor of seven as
compared to both FFN and LSSM rates. Otherwise the rates are in
relatively good comparison specially as the core shifts to higher
densities. As mentioned before the neutrino energy loss rates have
contributions both from electron capture and positron decay rates
(Eq. ~(\ref{nurate})). At $\rho Y_{e} [gcm^{-3}] =10$ and T$_{9} =
$1, the pn-QRPA calculated electron capture rates are in reasonable
comparison with the FFN and LSSM rates. On the other hand the
pn-QRPA calculated positron decay rates are bigger by roughly 8
orders of magnitude. The decay rates are very sensitive function of
available phase space ($= Q_{\beta} +E_{i} - E_{j}$). It is to be
noted that Brink's hypothesis is not assumed in the current
calculation. Brink's hypothesis states that GT strength distribution
on excited states is \textit{identical} to that from ground state,
shifted \textit{only} by the excitation energy of the state. In the
current pn-QRPA calculation all excited states are constructed in a
microscopic fashion as discussed earlier. This greatly increases the
reliability of calculated rates. Since the electron capture is the
dominant process the overall neutrino cooling rates is bigger only
by a factor of seven at $\rho Y_{e} [gcm^{-3}] =10$ and T$_{9} = $1.
As temperature increases the FFN and LSSM positron decay rates get
in better comparison with the pn-QRPA rates whereas the FFN and LSSM
electron capture rates surpass the pn-QRPA rates (due to a lower
placement of the centroid of the GT$_{+}$ strength distribution).
The reduced phase space at low temperatures is increased by finite
occupation probabilities of parent excitation energies at high
temperatures. At high temperatures the probability of occupation of
the parent excited states $({E_{i}})$ increases. FFN did not take
into effect the process of particle emission from excited states
(this process is accounted for in the present pn-QRPA calculation).
FFN's parent excitation energies $({E_{i}})$ are well above the
particle decay channel and partly contribute to the enhancement of
their weak rates at higher temperatures.

The comparison with the previous calculations improve at higher
stellar densities. The situation is depicted in Figure~\ref{figure5}
at stellar densities $\rho Y_{e} [gcm^{-3}] =10^{6}, 10^{7},
10^{8}$. At low temperatures the comparison is excellent. This is
roughly the region where weak interaction rates due to $^{56}$Ni is
considered to be most effective during the presupernova evolution of
massive stars. Core-collapse simulators might find it interesting to
note that all three calculations agree very nicely for the above
mentioned range of stellar temperatures and densities. As T$_{9}
\sim $10, the LSSM and FFN cooling rates become stronger for reasons
mentioned before. However the abundance of $^{56}$Ni also decreases
appreciably at high temperatures.

In high density regions the LSSM and FFN rates are bigger as
expected. The situation is depicted in Figure~\ref{figure6}. Here
one notes that the LSSM rates are bigger by as much as a factor of
five at $\rho Y_{e} [gcm^{-3}] =10^{9}$, 1 $\leq T_{9} \leq $ 3 . At
high stellar densities the weak rates are sensitive to the total GT
strength rather than its distribution details. The LSSM calculated
the total GT strength to be 10.1 \cite{Lan98} as compared to the
pn-QRPA value of 8.9 \cite{Nab08}. The larger total GT strength of
LSSM resulted in the enhancement of their rates at high densities.
The corresponding FFN rates are enhanced at most by a factor of
three at $\rho Y_{e} [gcm^{-3}] =10^{11}$ to an order of magnitude
at $\rho Y_{e} [gcm^{-3}] =10^{9}$.

\section{Conclusions}
The pn-QRPA theory was used to calculate the weak-interaction
mediated neutrino and antineutrino cooling rates due to $^{56}$Ni on
a detailed temperature-density grid point suitable for simulation
purposes.  At low temperatures and densities the pn-QRPA cooling
rates are enhanced. Otherwise the rates are in reasonable comparison
with previous calculations. For physical conditions considered to be
most effective for electron capture rates on $^{56}$Ni ($\rho Y_{e}
[gcm^{-3}] \sim 10^{7}$, 1 $\leq T_{9} \leq $ 5) the three
calculations are in very good agreement. Whereas for high stellar
temperatures and densities the LSSM and FFN cooling rates are much
bigger (up to an order of magnitude). However the abundance of
$^{56}$Ni decreases appreciably at high temperatures and densities.

According to Aufderheide and collaborators \cite{Auf94}, for $Y_{e}$
around 0.5, $^{56}$Ni is the most abundant nucleus having a mass
fraction of around 0.99. The mass fraction of most abundant nuclei
decreases appreciably as the $Y_{e}$ value decreases (e.g. it is of
the order of 10$^{-2}$ when $Y_{e}$ $\sim$ 0.46 and decreases by
another two orders of magnitude for still lower values of $Y_{e}$).
During the earlier phases of presupernova evolution, due to its high
abundance, electron capture on $^{56}$Ni is very important and the
rate of change of $Y_{e}$ is roughly around 25$\%$ alone due to
electron capture on $^{56}$Ni \cite{Auf94}. It is expected that the
neutrino cooling rates due to $^{56}$Ni might also have an effect on
the presupernova evolution of massive stars. It is expected that the
reported rates (see also \cite{Nab05,Nab08}) might contribute in the
fine-tuning of the $Y_{e}$ parameter during the various phases of
stellar evolution of massive stars. Core-collapse simulators are
suggested to check for possible interesting outcomes using the
reported neutrino cooling rates.

\ack The author wishes to acknowledge the support of research grant
provided by the Higher Education Commission, Pakistan  through the
HEC Project No. 20-1283.

\begin{figure}[htbp]
\begin{center}
\includegraphics[width=0.8\textwidth]{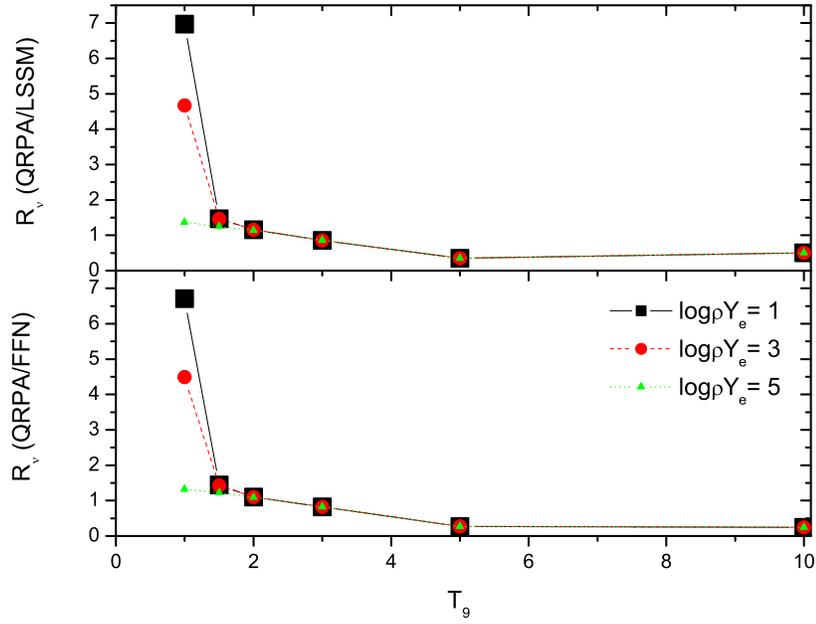}
\caption{(Color online) Ratios of pn-QRPA neutrino cooling rates to
those calculated using LSSM \cite{Lan00} (upper panel) and by FFN
\cite{Ful82} (lower panel) as function of stellar temperatures and
densities. T$_{9}$ gives the stellar temperature in units of
$10^{9}$ K. In the legend, log $\rho Y_{e}$ gives the log to base 10
of stellar density in units of $gcm^{-3}$.}\label{figure4}
\end{center}
\end{figure}
\begin{figure}[htbp]
\begin{center}
\includegraphics[width=0.8\textwidth]{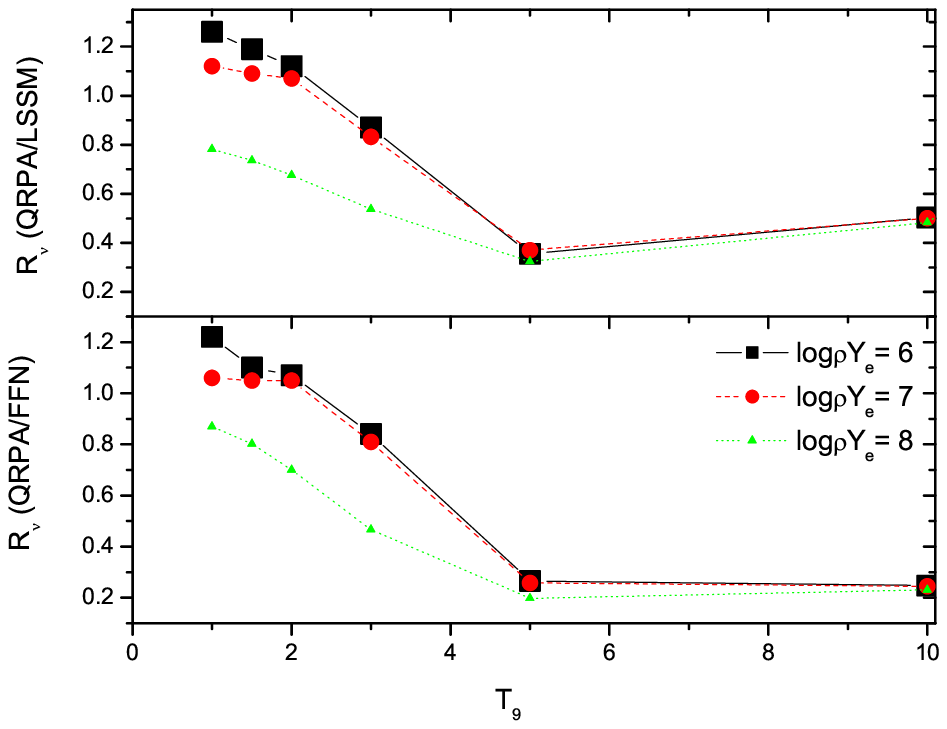}
\caption{(Color online) same as Figure~\ref{figure4}}\label{figure5}
\end{center}
\end{figure}
\begin{figure}[htbp]
\begin{center}
\includegraphics[width=0.8\textwidth]{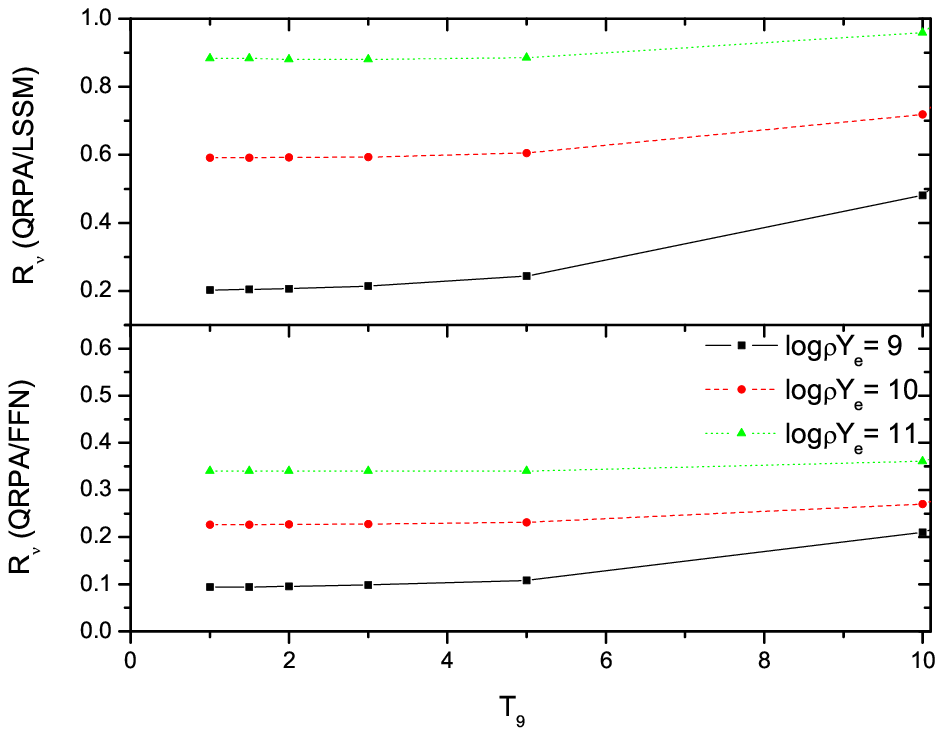}
\caption{(Color online) same as Figure~\ref{figure4}}\label{figure6}
\end{center}
\end{figure}

\clearpage\textbf{Table I:} Weak-interaction mediated neutrino and
antineutrino cooling rates due to $^{56}$Ni for selected densities
and temperatures in stellar matter. log$\rho Y_{e}$ has units of
$g/cm^{3}$, where $\rho$ is the baryon density and $Y_{e}$ is the
ratio of the lepton number to the baryon number. Temperatures
($T_{9}$) are given in units of $10^{9}$ K. The calculated Fermi
energy is denoted by $E_{f}$ and is given in units of MeV.
$\lambda_{\nu}$ ($\lambda_{\bar{\nu}}$) are the total neutrino
(antineutrino) cooling rates  as a result of $\beta^{+}$ decay and
electron capture ($\beta^{-}$ decay and positron capture) in units
of $MeV s^{-1}$. All calculated rates are tabulated in logarithmic
(to base 10) scale. In the table, -100 means that the rate is
smaller than 10$^{-100} MeV s^{-1}$.

\scriptsize{\begin{center}
\begin{tabular}{ccccc|ccccc|ccccc} \\ \hline
$log\rho Y_{e}$& $T_{9}$ & $E_{f}$& $\lambda_{\nu}$
&$\lambda_{\bar{\nu}}$& $log\rho Y_{e}$& $T_{9}$ & $E_{f}$&
$\lambda_{\nu}$ &$\lambda_{\bar{\nu}}$& $log\rho Y_{e}$& $T_{9}$
&$E_{f}$& $\lambda_{\nu}$ &$\lambda_{\bar{\nu}}$
\\\hline
0.5 &   0.50 &  0.065  & -6.917  &  -100  & 1.5  &   0.50  & 0.162  & -6.914  &  -100  &   2.5  &   0.50  &  0.261  & -6.887  &  -100   \\
0.5 &   1.00 &  0.000  & -6.826  &   -81.247  & 1.5  &   1.00  & 0.002  & -6.825  &   -81.254  &   2.5  &   1.00  &  0.015  & -6.811  &   -81.322   \\
0.5 &   1.50 &  0.000  & -6.152  &   -54.126  & 1.5  &   1.50  & 0.000  & -6.151  &   -54.126  &   2.5  &   1.50  &  0.002  & -6.147  &   -54.130   \\
0.5 &   2.00 &  0.000  & -5.389  &   -40.456  & 1.5  &   2.00  & 0.000  & -5.388  &   -40.455  &   2.5  &   2.00  &  0.000  & -5.387  &   -40.456   \\
0.5 &   2.50 &  0.000  & -4.782  &   -32.190  & 1.5  &   2.50  & 0.000  & -4.781  &   -32.190  &   2.5  &   2.50  &  0.000  & -4.780  &   -32.190   \\
0.5 &   3.00 &  0.000  & -4.290  &   -26.636  & 1.5  &   3.00  & 0.000  & -4.289  &   -26.636  &   2.5  &   3.00  &  0.000  & -4.288  &   -26.636   \\
0.5 &   3.50 &  0.000  & -3.873  &   -22.636  & 1.5  &   3.50  & 0.000  & -3.872  &   -22.635  &   2.5  &   3.50  &  0.000  & -3.872  &   -22.635   \\
0.5 &   4.00 &  0.000  & -3.506  &   -19.609  & 1.5  &   4.00  & 0.000  & -3.505  &   -19.608  &   2.5  &   4.00  &  0.000  & -3.505  &   -19.608   \\
0.5 &   4.50 &  0.000  & -3.167  &   -17.232  & 1.5  &   4.50  & 0.000  & -3.166  &   -17.231  &   2.5  &   4.50  &  0.000  & -3.166  &   -17.231   \\
0.5 &   5.00 &  0.000  & -2.843  &   -15.312  & 1.5  &   5.00  & 0.000  & -2.842  &   -15.310  &   2.5  &   5.00  &  0.000  & -2.842  &   -15.310   \\
0.5 &   5.50 &  0.000  & -2.529  &   -13.724  & 1.5  &   5.50  & 0.000  & -2.528  &   -13.722  &   2.5  &   5.50  &  0.000  & -2.528  &   -13.722   \\
0.5 &   6.00 &  0.000  & -2.223  &   -12.386  & 1.5  &   6.00  & 0.000  & -2.222  &   -12.384  &   2.5  &   6.00  &  0.000  & -2.222  &   -12.384   \\
0.5 &   6.50 &  0.000  & -1.928  &   -11.241  & 1.5  &   6.50  & 0.000  & -1.927  &   -11.239  &   2.5  &   6.50  &  0.000  & -1.927  &   -11.239   \\
0.5 &   7.00 &  0.000  & -1.645  &   -10.249  & 1.5  &   7.00  & 0.000  & -1.644  &   -10.246  &   2.5  &   7.00  &  0.000  & -1.644  &   -10.246   \\
0.5 &   7.50 &  0.000  & -1.376  &    -9.378  & 1.5  &   7.50  & 0.000  & -1.375  &    -9.375  &   2.5  &   7.50  &  0.000  & -1.375  &    -9.375   \\
0.5 &   8.00 &  0.000  & -1.121  &    -8.606  & 1.5  &   8.00  & 0.000  & -1.119  &    -8.604  &   2.5  &   8.00  &  0.000  & -1.119  &    -8.604   \\
0.5 &   8.50 &  0.000  & -0.879  &    -7.917  & 1.5  &   8.50  & 0.000  & -0.877  &    -7.915  &   2.5  &   8.50  &  0.000  & -0.877  &    -7.914   \\
0.5 &   9.00 &  0.000  & -0.650  &    -7.297  & 1.5  &   9.00  & 0.000  & -0.648  &    -7.294  &   2.5  &   9.00  &  0.000  & -0.648  &    -7.294   \\
0.5 &   9.50 &  0.000  & -0.433  &    -6.735  & 1.5  &   9.50  & 0.000  & -0.431  &    -6.732  &   2.5  &   9.50  &  0.000  & -0.431  &    -6.732   \\
0.5 &  10.00 &  0.000  & -0.227  &    -6.222  & 1.5  &  10.00  & 0.000  & -0.225  &    -6.220  &   2.5  &  10.00  &  0.000  & -0.225  &    -6.219   \\
0.5 &  20.00 &  0.000  &  2.433  &    -0.821  & 1.5  &  20.00  & 0.000  &  2.436  &    -0.817  &   2.5  &  20.00  &  0.000  &  2.436  &    -0.817   \\
0.5 &  30.00 &  0.000  &  3.787  &     1.457  & 1.5  &  30.00  & 0.000  &  3.791  &     1.462  &   2.5  &  30.00  &  0.000  &  3.791  &     1.463   \\
1.0 &   0.50 &  0.113  & -6.916  &  -100  & 2.0  &   0.50  & 0.212  & -6.908  &  -100  &   3.0  &   0.50  &  0.311  & -6.828  &  -100  \\
1.0 &   1.00 &  0.000  & -6.826  &   -81.249  & 2.0  &   1.00  & 0.005  & -6.822  &   -81.270  &   3.0  &   1.00  &  0.046  & -6.773  &   -81.476   \\
1.0 &   1.50 &  0.000  & -6.151  &   -54.126  & 2.0  &   1.50  & 0.000  & -6.150  &   -54.127  &   3.0  &   1.50  &  0.005  & -6.138  &   -54.141   \\
1.0 &   2.00 &  0.000  & -5.388  &   -40.455  & 2.0  &   2.00  & 0.000  & -5.388  &   -40.456  &   3.0  &   2.00  &  0.001  & -5.384  &   -40.459   \\
1.0 &   2.50 &  0.000  & -4.781  &   -32.190  & 2.0  &   2.50  & 0.000  & -4.781  &   -32.190  &   3.0  &   2.50  &  0.001  & -4.780  &   -32.191   \\
1.0 &   3.00 &  0.000  & -4.289  &   -26.636  & 2.0  &   3.00  & 0.000  & -4.288  &   -26.635  &   3.0  &   3.00  &  0.000  & -4.288  &   -26.636   \\
1.0 &   3.50 &  0.000  & -3.872  &   -22.635  & 2.0  &   3.50  & 0.000  & -3.872  &   -22.635  &   3.0  &   3.50  &  0.000  & -3.872  &   -22.635   \\
1.0 &   4.00 &  0.000  & -3.505  &   -19.608  & 2.0  &   4.00  & 0.000  & -3.505  &   -19.608  &   3.0  &   4.00  &  0.000  & -3.504  &   -19.608   \\
1.0 &   4.50 &  0.000  & -3.166  &   -17.231  & 2.0  &   4.50  & 0.000  & -3.166  &   -17.231  &   3.0  &   4.50  &  0.000  & -3.165  &   -17.231   \\
1.0 &   5.00 &  0.000  & -2.842  &   -15.310  & 2.0  &   5.00  & 0.000  & -2.842  &   -15.310  &   3.0  &   5.00  &  0.000  & -2.842  &   -15.310   \\
1.0 &   5.50 &  0.000  & -2.528  &   -13.723  & 2.0  &   5.50  & 0.000  & -2.528  &   -13.722  &   3.0  &   5.50  &  0.000  & -2.527  &   -13.722   \\
1.0 &   6.00 &  0.000  & -2.222  &   -12.385  & 2.0  &   6.00  & 0.000  & -2.222  &   -12.384  &   3.0  &   6.00  &  0.000  & -2.222  &   -12.384   \\
1.0 &   6.50 &  0.000  & -1.927  &   -11.240  & 2.0  &   6.50  & 0.000  & -1.927  &   -11.239  &   3.0  &   6.50  &  0.000  & -1.927  &   -11.239   \\
1.0 &   7.00 &  0.000  & -1.644  &   -10.247  & 2.0  &   7.00  & 0.000  & -1.644  &   -10.246  &   3.0  &   7.00  &  0.000  & -1.644  &   -10.246   \\
1.0 &   7.50 &  0.000  & -1.375  &    -9.376  & 2.0  &   7.50  & 0.000  & -1.375  &    -9.375  &   3.0  &   7.50  &  0.000  & -1.375  &    -9.375   \\
1.0 &   8.00 &  0.000  & -1.120  &    -8.605  & 2.0  &   8.00  & 0.000  & -1.119  &    -8.604  &   3.0  &   8.00  &  0.000  & -1.119  &    -8.604   \\
1.0 &   8.50 &  0.000  & -0.878  &    -7.915  & 2.0  &   8.50  & 0.000  & -0.877  &    -7.915  &   3.0  &   8.50  &  0.000  & -0.877  &    -7.914   \\
1.0 &   9.00 &  0.000  & -0.648  &    -7.295  & 2.0  &   9.00  & 0.000  & -0.648  &    -7.294  &   3.0  &   9.00  &  0.000  & -0.648  &    -7.294   \\
1.0 &   9.50 &  0.000  & -0.431  &    -6.733  & 2.0  &   9.50  & 0.000  & -0.431  &    -6.732  &   3.0  &   9.50  &  0.000  & -0.431  &    -6.732   \\
1.0 &  10.00 &  0.000  & -0.226  &    -6.220  & 2.0  &  10.00  & 0.000  & -0.225  &    -6.219  &   3.0  &  10.00  &  0.000  & -0.225  &    -6.219   \\
1.0 &  20.00 &  0.000  &  2.435  &    -0.818  & 2.0  &  20.00  & 0.000  &  2.436  &    -0.817  &   3.0  &  20.00  &  0.000  &  2.436  &    -0.816   \\
1.0 &  30.00 &  0.000  &  3.790  &     1.461  & 2.0  &  30.00  & 0.000  &  3.791  &     1.463  &   3.0  &  30.00  &  0.000  &  3.791  &     1.463   \\
\end{tabular}
\end{center}}
\scriptsize{\begin{center}
\begin{tabular}{ccccc|ccccc|ccccc} \\ \hline
$log\rho Y_{e}$& $T_{9}$ & $E_{f}$& $\lambda_{\nu}$
&$\lambda_{\bar{\nu}}$& $log\rho Y_{e}$& $T_{9}$ & $E_{f}$&
$\lambda_{\nu}$ &$\lambda_{\bar{\nu}}$& $log\rho Y_{e}$& $T_{9}$
&$E_{f}$& $\lambda_{\nu}$ &$\lambda_{\bar{\nu}}$
\\\hline
     3.5  &   0.50  & 0.361  &  -6.682  & -100  &      4.5  &   0.50  & 0.464  &  -6.009  & -100  &      5.5  &   0.50  & 0.598  &  -5.074  & -100  \\
     3.5  &   1.00  & 0.115  &  -6.643  & -81.824  &      4.5  &   1.00  & 0.309  &  -5.953  & -82.804  &      5.5  &   1.00  & 0.528  &  -4.982  & -83.910  \\
     3.5  &   1.50  & 0.015  &  -6.109  & -54.176  &      4.5  &   1.50  & 0.130  &  -5.768  & -54.561  &      5.5  &   1.50  & 0.423  &  -4.841  & -55.547  \\
     3.5  &   2.00  & 0.004  &  -5.377  & -40.466  &      4.5  &   2.00  & 0.044  &  -5.281  & -40.565  &      5.5  &   2.00  & 0.295  &  -4.666  & -41.200  \\
     3.5  &   2.50  & 0.002  &  -4.777  & -32.194  &      4.5  &   2.50  & 0.020  &  -4.741  & -32.230  &      5.5  &   2.50  & 0.180  &  -4.424  & -32.553  \\
     3.5  &   3.00  & 0.001  &  -4.287  & -26.637  &      4.5  &   3.00  & 0.011  &  -4.270  & -26.654  &      5.5  &   3.00  & 0.110  &  -4.107  & -26.819  \\
     3.5  &   3.50  & 0.001  &  -3.871  & -22.636  &      4.5  &   3.50  & 0.007  &  -3.862  & -22.645  &      5.5  &   3.50  & 0.072  &  -3.771  & -22.738  \\
     3.5  &   4.00  & 0.001  &  -3.504  & -19.608  &      4.5  &   4.00  & 0.005  &  -3.498  & -19.614  &      5.5  &   4.00  & 0.050  &  -3.443  & -19.671  \\
     3.5  &   4.50  & 0.000  &  -3.165  & -17.231  &      4.5  &   4.50  & 0.004  &  -3.162  & -17.235  &      5.5  &   4.50  & 0.038  &  -3.126  & -17.272  \\
     3.5  &   5.00  & 0.000  &  -2.842  & -15.310  &      4.5  &   5.00  & 0.003  &  -2.839  & -15.313  &      5.5  &   5.00  & 0.029  &  -2.816  & -15.339  \\
     3.5  &   5.50  & 0.000  &  -2.527  & -13.722  &      4.5  &   5.50  & 0.002  &  -2.526  & -13.724  &      5.5  &   5.50  & 0.023  &  -2.510  & -13.743  \\
     3.5  &   6.00  & 0.000  &  -2.222  & -12.384  &      4.5  &   6.00  & 0.002  &  -2.221  & -12.386  &      5.5  &   6.00  & 0.019  &  -2.209  & -12.400  \\
     3.5  &   6.50  & 0.000  &  -1.927  & -11.239  &      4.5  &   6.50  & 0.002  &  -1.926  & -11.240  &      5.5  &   6.50  & 0.016  &  -1.917  & -11.251  \\
     3.5  &   7.00  & 0.000  &  -1.644  & -10.246  &      4.5  &   7.00  & 0.001  &  -1.643  & -10.247  &      5.5  &   7.00  & 0.013  &  -1.637  & -10.256  \\
     3.5  &   7.50  & 0.000  &  -1.374  &  -9.375  &      4.5  &   7.50  & 0.001  &  -1.374  &  -9.376  &      5.5  &   7.50  & 0.012  &  -1.369  &  -9.383  \\
     3.5  &   8.00  & 0.000  &  -1.119  &  -8.604  &      4.5  &   8.00  & 0.001  &  -1.119  &  -8.604  &      5.5  &   8.00  & 0.010  &  -1.114  &  -8.610  \\
     3.5  &   8.50  & 0.000  &  -0.877  &  -7.914  &      4.5  &   8.50  & 0.001  &  -0.877  &  -7.915  &      5.5  &   8.50  & 0.009  &  -0.873  &  -7.919  \\
     3.5  &   9.00  & 0.000  &  -0.648  &  -7.294  &      4.5  &   9.00  & 0.001  &  -0.648  &  -7.294  &      5.5  &   9.00  & 0.008  &  -0.644  &  -7.298  \\
     3.5  &   9.50  & 0.000  &  -0.431  &  -6.732  &      4.5  &   9.50  & 0.001  &  -0.430  &  -6.732  &      5.5  &   9.50  & 0.007  &  -0.428  &  -6.735  \\
     3.5  &  10.00  & 0.000  &  -0.225  &  -6.219  &      4.5  &  10.00  & 0.001  &  -0.225  &  -6.219  &      5.5  &  10.00  & 0.006  &  -0.222  &  -6.222  \\
     3.5  &  20.00  & 0.000  &   2.436  &  -0.816  &      4.5  &  20.00  & 0.000  &   2.436  &  -0.816  &      5.5  &  20.00  & 0.001  &   2.437  &  -0.817  \\
     3.5  &  30.00  & 0.000  &   3.791  &   1.463  &      4.5  &  30.00  & 0.000  &   3.791  &   1.463  &      5.5  &  30.00  & 0.001  &   3.791  &   1.463  \\
     4.0  &   0.50  & 0.411  &  -6.403  & -100  &      5.0  &   0.50  & 0.522  &  -5.556  & -100  &      6.0  &   0.50  & 0.713  &  -4.537  & -100  \\
     4.0  &   1.00  & 0.209  &  -6.360  & -82.299  &      5.0  &   1.00  & 0.413  &  -5.483  & -83.330  &      6.0  &   1.00  & 0.672  &  -4.439  & -84.634  \\
     4.0  &   1.50  & 0.047  &  -6.017  & -54.283  &      5.0  &   1.50  & 0.265  &  -5.340  & -55.017  &      6.0  &   1.50  & 0.604  &  -4.304  & -56.154  \\
     4.0  &   2.00  & 0.014  &  -5.354  & -40.490  &      5.0  &   2.00  & 0.128  &  -5.074  & -40.779  &      6.0  &   2.00  & 0.512  &  -4.151  & -41.746  \\
     4.0  &   2.50  & 0.006  &  -4.768  & -32.202  &      5.0  &   2.50  & 0.062  &  -4.658  & -32.315  &      6.0  &   2.50  & 0.405  &  -3.986  & -33.006  \\
     4.0  &   3.00  & 0.004  &  -4.283  & -26.641  &      5.0  &   3.00  & 0.035  &  -4.230  & -26.695  &      6.0  &   3.00  & 0.299  &  -3.796  & -27.138  \\
     4.0  &   3.50  & 0.002  &  -3.869  & -22.638  &      5.0  &   3.50  & 0.023  &  -3.840  & -22.668  &      6.0  &   3.50  & 0.214  &  -3.570  & -22.943  \\
     4.0  &   4.00  & 0.002  &  -3.503  & -19.609  &      5.0  &   4.00  & 0.016  &  -3.485  & -19.628  &      6.0  &   4.00  & 0.156  &  -3.315  & -19.804  \\
     4.0  &   4.50  & 0.001  &  -3.164  & -17.232  &      5.0  &   4.50  & 0.012  &  -3.153  & -17.244  &      6.0  &   4.50  & 0.117  &  -3.043  & -17.362  \\
     4.0  &   5.00  & 0.001  &  -2.841  & -15.311  &      5.0  &   5.00  & 0.009  &  -2.834  & -15.319  &      6.0  &   5.00  & 0.091  &  -2.760  & -15.402  \\
     4.0  &   5.50  & 0.001  &  -2.527  & -13.723  &      5.0  &   5.50  & 0.007  &  -2.522  & -13.729  &      6.0  &   5.50  & 0.073  &  -2.472  & -13.789  \\
     4.0  &   6.00  & 0.001  &  -2.221  & -12.384  &      5.0  &   6.00  & 0.006  &  -2.218  & -12.389  &      6.0  &   6.00  & 0.060  &  -2.182  & -12.434  \\
     4.0  &   6.50  & 0.001  &  -1.926  & -11.239  &      5.0  &   6.50  & 0.005  &  -1.924  & -11.243  &      6.0  &   6.50  & 0.050  &  -1.897  & -11.278  \\
     4.0  &   7.00  & 0.000  &  -1.644  & -10.246  &      5.0  &   7.00  & 0.004  &  -1.642  & -10.249  &      6.0  &   7.00  & 0.042  &  -1.621  & -10.277  \\
     4.0  &   7.50  & 0.000  &  -1.374  &  -9.375  &      5.0  &   7.50  & 0.004  &  -1.373  &  -9.377  &      6.0  &   7.50  & 0.036  &  -1.356  &  -9.399  \\
     4.0  &   8.00  & 0.000  &  -1.119  &  -8.604  &      5.0  &   8.00  & 0.003  &  -1.117  &  -8.605  &      6.0  &   8.00  & 0.032  &  -1.104  &  -8.623  \\
     4.0  &   8.50  & 0.000  &  -0.877  &  -7.914  &      5.0  &   8.50  & 0.003  &  -0.876  &  -7.916  &      6.0  &   8.50  & 0.028  &  -0.864  &  -7.931  \\
     4.0  &   9.00  & 0.000  &  -0.648  &  -7.294  &      5.0  &   9.00  & 0.002  &  -0.647  &  -7.295  &      6.0  &   9.00  & 0.025  &  -0.637  &  -7.307  \\
     4.0  &   9.50  & 0.000  &  -0.431  &  -6.732  &      5.0  &   9.50  & 0.002  &  -0.430  &  -6.733  &      6.0  &   9.50  & 0.022  &  -0.421  &  -6.743  \\
     4.0  &  10.00  & 0.000  &  -0.225  &  -6.219  &      5.0  &  10.00  & 0.002  &  -0.224  &  -6.220  &      6.0  &  10.00  & 0.020  &  -0.217  &  -6.229  \\
     4.0  &  20.00  & 0.000  &   2.436  &  -0.816  &      5.0  &  20.00  & 0.000  &   2.436  &  -0.816  &      6.0  &  20.00  & 0.005  &   2.437  &  -0.817  \\
     4.0  &  30.00  & 0.000  &   3.791  &   1.463  &      5.0  &  30.00  & 0.000  &   3.791  &   1.463  &      6.0  &  30.00  & 0.002  &   3.792  &   1.463  \\
\end{tabular}
\end{center}}
\scriptsize{\begin{center}
\begin{tabular}{ccccc|ccccc|ccccc} \\ \hline
$log\rho Y_{e}$& $T_{9}$ & $E_{f}$& $\lambda_{\nu}$
&$\lambda_{\bar{\nu}}$& $log\rho Y_{e}$& $T_{9}$ & $E_{f}$&
$\lambda_{\nu}$ &$\lambda_{\bar{\nu}}$& $log\rho Y_{e}$& $T_{9}$
&$E_{f}$& $\lambda_{\nu}$ &$\lambda_{\bar{\nu}}$
\\\hline
     6.5  &   0.50  & 0.905  &  -3.883  & -100  &      7.5  &   0.50  & 1.705  &  -2.203  & -100  &      8.5  &   0.50  & 3.547  &  -0.270  & -100   \\
     6.5  &   1.00  & 0.880  &  -3.807  & -85.680  &      7.5  &   1.00  & 1.693  &  -2.178  & -89.781  &      8.5  &   1.00  & 3.542  &  -0.264  & -99.099   \\
     6.5  &   1.50  & 0.837  &  -3.697  & -56.937  &      7.5  &   1.50  & 1.675  &  -2.139  & -59.752  &      8.5  &   1.50  & 3.534  &  -0.253  & -65.998   \\
     6.5  &   2.00  & 0.777  &  -3.570  & -42.413  &      7.5  &   2.00  & 1.648  &  -2.087  & -44.609  &      8.5  &   2.00  & 3.521  &  -0.238  & -49.329   \\
     6.5  &   2.50  & 0.701  &  -3.434  & -33.602  &      7.5  &   2.50  & 1.614  &  -2.026  & -35.444  &      8.5  &   2.50  & 3.506  &  -0.218  & -39.257   \\
     6.5  &   3.00  & 0.612  &  -3.295  & -27.664  &      7.5  &   3.00  & 1.573  &  -1.956  & -29.278  &      8.5  &   3.00  & 3.487  &  -0.192  & -32.493   \\
     6.5  &   3.50  & 0.517  &  -3.149  & -23.379  &      7.5  &   3.50  & 1.524  &  -1.881  & -24.830  &      8.5  &   3.50  & 3.464  &  -0.161  & -27.623   \\
     6.5  &   4.00  & 0.424  &  -2.989  & -20.142  &      7.5  &   4.00  & 1.468  &  -1.800  & -21.457  &      8.5  &   4.00  & 3.438  &  -0.122  & -23.939   \\
     6.5  &   4.50  & 0.343  &  -2.804  & -17.614  &      7.5  &   4.50  & 1.405  &  -1.710  & -18.804  &      8.5  &   4.50  & 3.408  &  -0.076  & -21.048   \\
     6.5  &   5.00  & 0.277  &  -2.591  & -15.589  &      7.5  &   5.00  & 1.336  &  -1.610  & -16.657  &      8.5  &   5.00  & 3.375  &  -0.020  & -18.712   \\
     6.5  &   5.50  & 0.226  &  -2.352  & -13.929  &      7.5  &   5.50  & 1.262  &  -1.493  & -14.878  &      8.5  &   5.50  & 3.339  &   0.047  & -16.781   \\
     6.5  &   6.00  & 0.187  &  -2.095  & -12.541  &      7.5  &   6.00  & 1.183  &  -1.358  & -13.377  &      8.5  &   6.00  & 3.299  &   0.125  & -15.155   \\
     6.5  &   6.50  & 0.157  &  -1.832  & -11.361  &      7.5  &   6.50  & 1.101  &  -1.204  & -12.093  &      8.5  &   6.50  & 3.256  &   0.215  & -13.763   \\
     6.5  &   7.00  & 0.134  &  -1.571  & -10.342  &      7.5  &   7.00  & 1.018  &  -1.035  & -10.979  &      8.5  &   7.00  & 3.209  &   0.317  & -12.557   \\
     6.5  &   7.50  & 0.115  &  -1.316  &  -9.452  &      7.5  &   7.50  & 0.937  &  -0.857  & -10.004  &      8.5  &   7.50  & 3.159  &   0.428  & -11.498   \\
     6.5  &   8.00  & 0.100  &  -1.071  &  -8.666  &      7.5  &   8.00  & 0.858  &  -0.675  &  -9.144  &      8.5  &   8.00  & 3.106  &   0.545  & -10.560   \\
     6.5  &   8.50  & 0.088  &  -0.836  &  -7.966  &      7.5  &   8.50  & 0.783  &  -0.494  &  -8.378  &      8.5  &   8.50  & 3.050  &   0.666  &  -9.722   \\
     6.5  &   9.00  & 0.078  &  -0.613  &  -7.337  &      7.5  &   9.00  & 0.714  &  -0.315  &  -7.693  &      8.5  &   9.00  & 2.990  &   0.789  &  -8.968   \\
     6.5  &   9.50  & 0.069  &  -0.401  &  -6.768  &      7.5  &   9.50  & 0.651  &  -0.140  &  -7.077  &      8.5  &   9.50  & 2.928  &   0.910  &  -8.285   \\
     6.5  &  10.00  & 0.062  &  -0.199  &  -6.250  &      7.5  &  10.00  & 0.593  &   0.030  &  -6.518  &      8.5  &  10.00  & 2.863  &   1.030  &  -7.662   \\
     6.5  &  20.00  & 0.015  &   2.440  &  -0.820  &      7.5  &  20.00  & 0.150  &   2.473  &  -0.854  &      8.5  &  20.00  & 1.402  &   2.777  &  -1.170   \\
     6.5  &  30.00  & 0.007  &   3.792  &   1.462  &      7.5  &  30.00  & 0.066  &   3.802  &   1.452  &      8.5  &  30.00  & 0.656  &   3.900  &   1.353   \\
     7.0  &   0.50  & 1.217  &  -3.094  & -100  &      8.0  &   0.50  & 2.444  &  -1.253  & -100  &      9.0  &   0.50  & 5.179  &   0.825  & -100   \\
     7.0  &   1.00  & 1.200  &  -3.048  & -87.296  &      8.0  &   1.00  & 2.437  &  -1.241  & -93.527  &      9.0  &   1.00  & 5.176  &   0.832  & -100   \\
     7.0  &   1.50  & 1.173  &  -2.977  & -58.065  &      8.0  &   1.50  & 2.424  &  -1.221  & -62.269  &      9.0  &   1.50  & 5.170  &   0.844  & -71.496   \\
     7.0  &   2.00  & 1.133  &  -2.887  & -43.311  &      8.0  &   2.00  & 2.406  &  -1.194  & -46.518  &      9.0  &   2.00  & 5.162  &   0.862  & -53.462   \\
     7.0  &   2.50  & 1.083  &  -2.788  & -34.372  &      8.0  &   2.50  & 2.383  &  -1.160  & -36.994  &      9.0  &   2.50  & 5.151  &   0.884  & -42.574   \\
     7.0  &   3.00  & 1.021  &  -2.682  & -28.351  &      8.0  &   3.00  & 2.355  &  -1.120  & -30.591  &      9.0  &   3.00  & 5.138  &   0.911  & -35.267   \\
     7.0  &   3.50  & 0.950  &  -2.573  & -24.003  &      8.0  &   3.50  & 2.322  &  -1.074  & -25.978  &      9.0  &   3.50  & 5.122  &   0.942  & -30.011   \\
     7.0  &   4.00  & 0.871  &  -2.459  & -20.704  &      8.0  &   4.00  & 2.283  &  -1.020  & -22.484  &      9.0  &   4.00  & 5.105  &   0.979  & -26.039   \\
     7.0  &   4.50  & 0.785  &  -2.337  & -18.110  &      8.0  &   4.50  & 2.240  &  -0.959  & -19.739  &      9.0  &   4.50  & 5.085  &   1.020  & -22.925   \\
     7.0  &   5.00  & 0.698  &  -2.197  & -16.014  &      8.0  &   5.00  & 2.192  &  -0.886  & -17.520  &      9.0  &   5.00  & 5.062  &   1.065  & -20.412   \\
     7.0  &   5.50  & 0.613  &  -2.034  & -14.284  &      8.0  &   5.50  & 2.139  &  -0.799  & -15.682  &      9.0  &   5.50  & 5.037  &   1.114  & -18.338   \\
     7.0  &   6.00  & 0.534  &  -1.846  & -12.833  &      8.0  &   6.00  & 2.081  &  -0.696  & -14.132  &      9.0  &   6.00  & 5.010  &   1.169  & -16.592   \\
     7.0  &   6.50  & 0.465  &  -1.636  & -11.599  &      8.0  &   6.50  & 2.019  &  -0.577  & -12.804  &      9.0  &   6.50  & 4.980  &   1.229  & -15.101   \\
     7.0  &   7.00  & 0.404  &  -1.414  & -10.537  &      8.0  &   7.00  & 1.952  &  -0.444  & -11.652  &      9.0  &   7.00  & 4.948  &   1.296  & -13.809   \\
     7.0  &   7.50  & 0.353  &  -1.189  &  -9.612  &      8.0  &   7.50  & 1.882  &  -0.301  & -10.640  &      9.0  &   7.50  & 4.914  &   1.368  & -12.677   \\
     7.0  &   8.00  & 0.310  &  -0.965  &  -8.799  &      8.0  &   8.00  & 1.808  &  -0.153  &  -9.743  &      9.0  &   8.00  & 4.878  &   1.446  & -11.676   \\
     7.0  &   8.50  & 0.274  &  -0.748  &  -8.077  &      8.0  &   8.50  & 1.732  &  -0.003  &  -8.941  &      9.0  &   8.50  & 4.839  &   1.528  & -10.783   \\
     7.0  &   9.00  & 0.244  &  -0.537  &  -7.430  &      8.0  &   9.00  & 1.653  &   0.145  &  -8.220  &      9.0  &   9.00  & 4.797  &   1.615  &  -9.980   \\
     7.0  &   9.50  & 0.218  &  -0.335  &  -6.847  &      8.0  &   9.50  & 1.574  &   0.290  &  -7.566  &      9.0  &   9.50  & 4.754  &   1.705  &  -9.253   \\
     7.0  &  10.00  & 0.196  &  -0.142  &  -6.318  &      8.0  &  10.00  & 1.493  &   0.430  &  -6.972  &      9.0  &  10.00  & 4.708  &   1.796  &  -8.592   \\
     7.0  &  20.00  & 0.047  &   2.448  &  -0.828  &      8.0  &  20.00  & 0.470  &   2.551  &  -0.935  &      9.0  &  20.00  & 3.390  &   3.248  &  -1.670   \\
     7.0  &  30.00  & 0.021  &   3.795  &   1.460  &      8.0  &  30.00  & 0.209  &   3.826  &   1.428  &      9.0  &  30.00  & 1.973  &   4.115  &   1.132   \\
\end{tabular}
\end{center}}
\scriptsize{\begin{center}
\begin{tabular}{ccccc|ccccc} \\ \hline
$log\rho Y_{e}$& $T_{9}$ & $E_{f}$& $\lambda_{\nu}$
&$\lambda_{\bar{\nu}}$& $log\rho Y_{e}$& $T_{9}$ & $E_{f}$&
$\lambda_{\nu}$ &$\lambda_{\bar{\nu}}$
\\\hline
     9.5  &   0.50  & 7.583  &   2.296  & -100  &     10.5  &   0.50  & 16.310  &   4.902  & -100   \\
     9.5  &   1.00  & 7.581  &   2.299  & -100  &     10.5  &   1.00  & 16.309  &   4.902  & -100   \\
     9.5  &   1.50  & 7.577  &   2.305  & -79.583  &     10.5  &   1.50  & 16.307  &   4.903  & -100   \\
     9.5  &   2.00  & 7.571  &   2.313  & -59.534  &     10.5  &   2.00  & 16.304  &   4.904  & -81.542   \\
     9.5  &   2.50  & 7.564  &   2.323  & -47.438  &     10.5  &   2.50  & 16.301  &   4.905  & -65.052   \\
     9.5  &   3.00  & 7.555  &   2.335  & -39.328  &     10.5  &   3.00  & 16.297  &   4.907  & -54.014   \\
     9.5  &   3.50  & 7.545  &   2.349  & -33.499  &     10.5  &   3.50  & 16.292  &   4.908  & -46.095   \\
     9.5  &   4.00  & 7.532  &   2.365  & -29.098  &     10.5  &   4.00  & 16.286  &   4.910  & -40.128   \\
     9.5  &   4.50  & 7.519  &   2.383  & -25.651  &     10.5  &   4.50  & 16.280  &   4.912  & -35.464   \\
     9.5  &   5.00  & 7.503  &   2.403  & -22.873  &     10.5  &   5.00  & 16.273  &   4.915  & -31.713   \\
     9.5  &   5.50  & 7.486  &   2.424  & -20.582  &     10.5  &   5.50  & 16.265  &   4.918  & -28.626   \\
     9.5  &   6.00  & 7.468  &   2.448  & -18.657  &     10.5  &   6.00  & 16.256  &   4.921  & -26.039   \\
     9.5  &   6.50  & 7.448  &   2.475  & -17.014  &     10.5  &   6.50  & 16.247  &   4.925  & -23.837   \\
     9.5  &   7.00  & 7.426  &   2.504  & -15.593  &     10.5  &   7.00  & 16.237  &   4.930  & -21.936   \\
     9.5  &   7.50  & 7.403  &   2.537  & -14.349  &     10.5  &   7.50  & 16.226  &   4.936  & -20.279   \\
     9.5  &   8.00  & 7.378  &   2.574  & -13.251  &     10.5  &   8.00  & 16.214  &   4.943  & -18.818   \\
     9.5  &   8.50  & 7.351  &   2.615  & -12.273  &     10.5  &   8.50  & 16.202  &   4.952  & -17.521   \\
     9.5  &   9.00  & 7.323  &   2.661  & -11.394  &     10.5  &   9.00  & 16.189  &   4.963  & -16.359   \\
     9.5  &   9.50  & 7.293  &   2.711  & -10.600  &     10.5  &   9.50  & 16.175  &   4.977  & -15.313   \\
     9.5  &  10.00  & 7.261  &   2.764  &  -9.879  &     10.5  &  10.00  & 16.160  &   4.992  & -14.364   \\
     9.5  &  20.00  & 6.307  &   3.901  &  -2.405  &     10.5  &  20.00  & 15.711  &   5.593  &  -4.775   \\
     9.5  &  30.00  & 4.859  &   4.580  &   0.647  &     10.5  &  30.00  & 14.965  &   6.020  &  -1.051   \\
    10.0  &   0.50  & 11.118  &   3.677  & -100  &     11.0  &   0.50  & 23.934  &   6.024  & -100   \\
    10.0  &   1.00  & 11.116  &   3.679  & -100  &     11.0  &   1.00  & 23.933  &   6.024  & -100   \\
    10.0  &   1.50  & 11.113  &   3.680  & -91.466  &     11.0  &   1.50  & 23.932  &   6.024  & -100   \\
    10.0  &   2.00  & 11.110  &   3.683  & -68.451  &     11.0  &   2.00  & 23.930  &   6.024  & -100   \\
    10.0  &   2.50  & 11.105  &   3.686  & -54.576  &     11.0  &   2.50  & 23.928  &   6.025  & -80.427   \\
    10.0  &   3.00  & 11.099  &   3.690  & -45.281  &     11.0  &   3.00  & 23.925  &   6.025  & -66.829   \\
    10.0  &   3.50  & 11.091  &   3.695  & -38.606  &     11.0  &   3.50  & 23.922  &   6.026  & -57.081   \\
    10.0  &   4.00  & 11.083  &   3.700  & -33.572  &     11.0  &   4.00  & 23.918  &   6.027  & -49.743   \\
    10.0  &   4.50  & 11.074  &   3.706  & -29.633  &     11.0  &   4.50  & 23.913  &   6.028  & -44.013   \\
    10.0  &   5.00  & 11.063  &   3.713  & -26.462  &     11.0  &   5.00  & 23.908  &   6.029  & -39.409   \\
    10.0  &   5.50  & 11.052  &   3.721  & -23.849  &     11.0  &   5.50  & 23.903  &   6.030  & -35.626   \\
    10.0  &   6.00  & 11.039  &   3.730  & -21.657  &     11.0  &   6.00  & 23.897  &   6.031  & -32.458   \\
    10.0  &   6.50  & 11.025  &   3.739  & -19.788  &     11.0  &   6.50  & 23.891  &   6.033  & -29.763   \\
    10.0  &   7.00  & 11.011  &   3.751  & -18.174  &     11.0  &   7.00  & 23.884  &   6.036  & -27.442   \\
    10.0  &   7.50  & 10.995  &   3.764  & -16.763  &     11.0  &   7.50  & 23.877  &   6.039  & -25.420   \\
    10.0  &   8.00  & 10.978  &   3.779  & -15.519  &     11.0  &   8.00  & 23.869  &   6.043  & -23.641   \\
    10.0  &   8.50  & 10.959  &   3.797  & -14.412  &     11.0  &   8.50  & 23.860  &   6.048  & -22.062   \\
    10.0  &   9.00  & 10.940  &   3.818  & -13.420  &     11.0  &   9.00  & 23.851  &   6.056  & -20.650   \\
    10.0  &   9.50  & 10.920  &   3.843  & -12.525  &     11.0  &   9.50  & 23.842  &   6.064  & -19.380   \\
    10.0  &  10.00  & 10.898  &   3.870  & -11.712  &     11.0  &  10.00  & 23.832  &   6.075  & -18.230   \\
    10.0  &  20.00  & 10.241  &   4.690  &  -3.397  &     11.0  &  20.00  & 23.526  &   6.557  &  -6.744   \\
    10.0  &  30.00  & 9.163  &   5.236  &  -0.076  &     11.0  &  30.00  & 23.016  &   6.897  &  -2.403   \\
\end{tabular}
\end{center}}

\section*{References}
\begin{thebibliography}{99}
\bibitem{Baa34}Baade W and Zwicky F, 1934 {\it Proc. N. A. S. }{\bf20}
254; {\it Proc. N. A. S. }{\bf20} 259.
\bibitem{Col66}Colgate S A and White R, 1966 {\it Astrophys. J.} {\bf143} 626.
\bibitem{Arn67}Arnett W D, 1967 {\it Canadian J. Phys.} {\bf45} 1621.
\bibitem{Bet85} Bethe H A and Wilson J R, 1985 {\it Astrophys. J.}{\bf 295}
14.
\bibitem{Bur03}Buras R, Rampp M, Janka H-T and Kifonidis K, 2003
{\it Phys. Rev. Lett.} {\bf90}  241101.
\bibitem{Buras06}Buras R, Janka H-T, Rampp M and Kifonidis K, 2006
{\it Astron. Astrophys.} {\bf457}  281.
\bibitem{Bur06} Burrows A, Livne E, Dessart L and Ott C D, 2006
{\it Astrophys. J.} {\bf645} 534.
\bibitem{Woo07} Woosley S E and Heger A, 2007 {\it Phys. Rep.}{\bf442} 269.
\bibitem{Auf94} Aufderheide M B, Fushiki I, Woosley S E, Stanford E
and Hartmann D H, 1994 {\it Astrophys. J. Suppl. Ser} {\bf 91} 389.
\bibitem{Heg01} Heger A, Woosley S E, Mart\'{i}nez-Pinedo G and
Langanke K, 2001 {\it Astrophys. J.} {\bf 560} 307.
\bibitem{Nab05} Nabi J-Un and Rahman M-Ur, 2005 {\it Phys. Lett. B} {\bf612} 190.
\bibitem{Nab08}Nabi J-Un , Rahman M-Ur, and Sajjad M, 2008 {\it Acta. Phys. Polon. B} {\bf39} 651.
\bibitem{Jan07} Janka H-T, Langanke K, Marek A, Mart\'{i}nez-Pinedo
G and M\"{o}ller B, 2007 {\it Phys. Rep.}{\bf442} 38.
\bibitem{Yos88} Yost G P {\it et al.} (Particle Data Group), 1988 {\it Phys. Lett. B}
{\bf 204} 1.
\bibitem{Rod06} Rodin V, Faessler A, Simkovic F and Vogel P, 2006 {\it Czech. J.
Phys.} {\bf 56} 495.
\bibitem{Gov71}Gove N B and Martin M J, 1971 {\it At. Data Nucl. Data Tables} {\bf 10} 205.
\bibitem{Nab04} Nabi J-Un and Klapdor-Kleingrothaus H V 2004 {\it At. Data Nucl. Data Tables}
 {\bf 88} 237.
\bibitem{Mut92} Muto K, Bender E, Oda T and Klapdor H V, 1992 {\it
Zeit. Phys. A}{\bf341} 407.
\bibitem{Mut89} Muto K, Bender E and Klapdor H V, 1989 {\it
Zeit. Phys. A}{\bf333} 125.
\bibitem{Ful82} Fuller G M, Fowler W A and Newman M J, 1980 {\it Astrophys. J. Suppl.}
{\bf 42} 447; Fuller G M, Fowler W A and Newman M J, 1982 {\it
Astrophys. J. Suppl.} {\bf 48} 279; Fuller G M, Fowler W A and
Newman M J, 1982 {\it Astrophys. J.} {\bf 252} 715; Fuller G M,
Fowler W A and Newman M J, 1985 {\it Astrophys. J.} {\bf 293} 1.
\bibitem{Lan98} Langanke K and Mart\'{i}nez-Pinedo G, 1998 {\it Phys. Lett. B} {\bf436} 19.
\bibitem{Lan00} Langanke K and Mart\'{i}nez-Pinedo G, 2000 {\it Nucl. Phys.}{\bf A673}
481.
\bibitem{Wea78}Weaver T A, Zimmerman G B and Woosley S E, 1978 {\it Astrophys. J.}
{\bf225} 1021.
\end{thebibliography}
\end{document}